\PassOptionsToPackage{table}{xcolor}
\documentclass{IOS-Book-Article}

\usepackage{mathptmx}
\usepackage{soul}\setuldepth{article}
\usepackage{xspace}
\usepackage{float}
\usepackage{amssymb}
\usepackage{amsmath}
\usepackage{booktabs}
\usepackage{rotating}
\usepackage{arydshln}
\usepackage{multirow}
\usepackage{fontawesome5}
\usepackage{pifont}
\usepackage{makecell}
\usepackage[dvipsnames]{xcolor}
\usepackage{graphicx} 
\usepackage{svg}
\usepackage{subfigure}
\usepackage{bm}
\usepackage{pdfpages}
\usepackage{tcolorbox}
\usepackage{hyperref}
\hypersetup{colorlinks,allcolors=black}

\newcommand{\DSgaz}{DS:Gazetteer\xspace}
\newcommand{\DSqald}{DS:QALD9$_{\textrm{RDF/XML}}$\xspace}
\newcommand{\DSgn}{DS:GeoNames\xspace}
\newcommand{\DSgncountry}{DS:GeoNames$_{\textrm{country}}$\xspace}
\newcommand{\best}[1]{\underline{\textbf{#1}}}

\newcommand{\sortTopDown}{$\left( \downarrow^1_{k_{\textrm{ranker}}} \right)$\xspace}
\newcommand{\sortDownTop}{$\left( \downarrow^{k_{\textrm{ranker}}}_1 \right)$\xspace}
\newcommand{\symbCZero}{$\varnothing$}
\newcommand{\symbCOne}{\faComments}
\newcommand{\symbCTwoA}{\faGlobeAsia\hspace*{.5ex} $+$ \faComments\xspace}
\newcommand{\symbCTwoB}{\faGlobeAsia \hspace*{.1ex} \faCity\hspace*{.5ex} $+$ \faComments\xspace}
\newcommand{\symbCTwoC}{\faGlobeAsia \hspace*{.1ex} \faComments\xspace}

\def\hb{\hbox to 11.5 cm{}}

\begin{document}

\pagestyle{headings}
\def\thepage{}
\begin{frontmatter}              

\title{Identifying Origins of Place Names \textit{via} Retrieval Augmented Generation}


\author[A]{\fnms{Alexis} \snm{HORDE VO}\orcid{0009-0004-3324-3380}%
\thanks{Corresponding Author: Alexis HORDE VO, alexis.horde.vo@student.rmit.edu.au}},
\author[A]{\fnms{Matt} \snm{DUCKHAM}\orcid{0000-0002-7249-6709}},
\author[A]{\fnms{Estrid} \snm{HE}\orcid{0000-0002-8994-9532}}
and
\author[B]{\fnms{Rafe} \snm{BENLI}}

\address[A]{School of Computing Technologies, RMIT University, Melbourne, Australia}
\address[B]{Geographic Names Victoria, Victoria State Government, Australia}

\begin{abstract}
Who is the \textit{Batman} behind ``Batman Street'' in Melbourne? Understanding the historical, cultural, and societal narratives behind place names can reveal the rich context that has shaped a community. Although place names serve as essential spatial references in gazetteers, they often lack information about place name origins. Enriching these place names in today's gazetteers is a time-consuming, manual process that requires extensive exploration of a vast archive of documents and text sources. Recent advances in natural language processing and language models (LMs) hold the promise of significant automation of identifying place name origins due to their powerful capability to exploit the semantics of the stored documents. This chapter presents a retrieval augmented generation pipeline designed to search for place name origins over a broad knowledge base, DBpedia. Given a spatial query, our approach first extracts sub-graphs that may contain knowledge relevant to the query; then ranks the extracted sub-graphs to generate the final answer to the query using fine-tuned LM-based models (i.e., ColBERTv2 and Llama2). Our results highlight the key challenges facing automated retrieval of place name origins, especially the tendency of language models to under-use the spatial information contained in texts as a discriminating factor. Our approach also frames the wider implications for geographic information retrieval using retrieval augmented generation. 
\end{abstract}

\end{frontmatter}
\markboth{Horde Vo, Duckham, He, and Benli\hb}{Identifying Origins of Place Names Via Retrieval Augmented Generation\hb}
\section{Introduction}

Although place names are officially recorded by place naming authorities around the world, gazetteers commonly lack information regarding place name \textit{origins}. To understand the history behind place names, one must typically consult external sources, such as historical archives or web pages. Understanding that history is becoming increasingly important to communities, for example, in better reflecting the contributions of marginalized groups, such as women or Indigenous people, to a place. To increase diversity with commemorative places in Victoria (Australia), for example, gender equality policies face questions such as: what are the existing places named in honor of women? This question implies the detection of such place names as well as the generation of relevant arguments; in other words, what is the origin of a place name? Place names may refer to multiple objects (streets, buildings, cities or even natural features) but streets are most frequently encountered in daily life, e.g. when addressing a parcel or locating a restaurant near our workplaces. Accordingly, there is a growing need for automated tools that can efficiently retrieve relevant information about the origins of a given place name. With its rich available archives, this chapter evaluates these automated tools within the streets of Melbourne, the historic center of Victoria.

Developing such automated tools is challenging for at least two reasons. First, spatial queries can be highly ambiguous. For example, place names in Australia often consist of a single identifying word rather than a whole name (e.g., \textit{Batman} Street in Melbourne, contrary to Avenue \textit{Simone Veil} in Nice, France). To deal with ambiguity, it is often necessary to clarify the spatial context for a place name, such as the city, state, and country; the neighboring streets and the neighborhood. Our expectations about the most likely name origins vary spatially, for example, depending on whether the name appears in Melbourne, Victoria (where John Batman is a well-known historical figure who played a role in the founding of Melbourne, as well as in numerous massacres of Indigenous Australians) or in Los Angeles, California (where the comic-book character might be more relevant to the spatial context). 


Second, place name origins often fall within the domain of ``long-tail knowledge'': discovering origins may rely on piecing together multiple low-frequency but salient occurrences in a knowledge base. For example, in a knowledge base such as DBpedia the name ``Batman'' is likely to appear much more frequently in association with the popular superhero than the historical figure. Identifying such long-tail instances often requires a chain of sophisticated reasoning to uncover correct answers.

\begin{figure}
\centering
      \includegraphics[width=\textwidth]{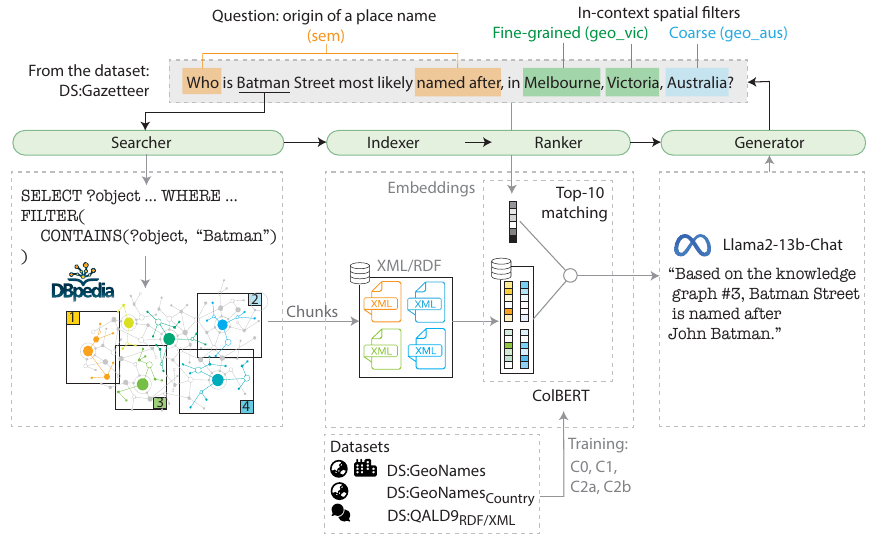}
\caption{An overview of our approach. }
\label{fig:general_schema}
\end{figure}

In this chapter, we develop a geographic information retrieval (GIR) system to automatically and effectively identify the origins of place names. We formulate the problem as a retrieval augmented generation task, allowing us to combine the strengths of traditional information retrieval and the generative capabilities of advanced language models (LMs). In our approach, relevant data is first retrieved from external knowledge sources (i.e., DBpedia), and then used to guide the generation of responses by pre-trained LMs (i.e., ColBERTv2 \cite{santhanamColBERTv2EffectiveEfficient2022} and Llama2 \cite{touvronLlamaOpenFoundation2023}). Consequently, the presented system is capable of processing user queries formulated in natural language and providing responses that are both contextually accurate and linguistically coherent. In order to enhance the spatial understanding of LMs, we inject spatial knowledge to ColBERTv2 by fine-tuning it on a dataset crafted from GeoNames. Figure~\ref{fig:general_schema} presents an overview of our approach: the \textit{searcher} retrieves relevant data from DBpedia, the \textit{indexer} and the \textit{ranker} filters the retrieved data via a fine-tuned ColBERTv2, and the \textit{generator} produces the final answer via Llama2. 

In this study of how best to leverage LMs for GIR, we focus in particular on two key questions:

\paragraph*{Can traditional LMs adequately reflect spatial relationships and geographic context?} We perform controlled experiments to evaluate various components of our model, such as semantic and spatial understanding. When specially tuned to spatial understanding, LMs become more limited in general semantic understanding. Further, these models do not reliably capture spatial containment, nor prioritize spatially proximal objects. Further, introducing in-context spatial filters only helps to \textit{partially orient} the predictions.     
\paragraph*{Can the use of external knowledge bases increase response accuracy?} By providing a context, the generator is driven to produce an output that relates more strongly to the provided ground knowledge, rather than fabricating (i.e., so-called ``hallucination'') nonexistent objects. However, we still observe ``mirages'': real but spatially distal objects that appear closer in terms of embeddings, and by consequence, acquire greater relevance for language models. Finally, the use of external knowledge restricts the solution space in a way that potentially increases the chance of correct answers; nonetheless, the models still need to learn how to navigate in this space.

\section{Background and related work}

\paragraph{Scope of information retrieval over knowledge bases.}Information retrieval aims to find ``(\textit{query, answer})'' pairs. In knowledge bases, data can have an unstructured format with text; a semi-structured format with tables; or a structured format with knowledge graphs \cite{purvesGeographicInformationRetrieval}. With the latter, knowledge graphs store information in atomic triplets ``(\textit{subject, predicate, object})'' and respect ontologies for semantic interoperability aspects. 
For unstructured data, conventional approaches rely on pre-trained language models that capture and compare dense representations with text embeddings \cite{lewisRetrievalAugmentedGenerationKnowledgeIntensive2021}. Querying semi-structured data typically implies converting the query to SQL. For structured data, current approaches either seek to convert the query to SPARQL or to find similarities with graph embeddings \cite{lanSurveyComplexKnowledge2021}. 

\paragraph*{Modular architectures for information retrieval architectures.} DBpedia is a widely used semantic knowledge base. However, it lacks consistency, mixing structured and unstructured data, such as paragraphs in natural language. As a result, reasoning and generalization with such knowledge bases is challenging. To tackle the problem, current architectures for information retrieval use multiple modules, such as an indexer to optimize the knowledge base for queries and a ranker to find the best matches \cite{gaoRetrievalAugmentedGenerationLarge2024}. Neural rankers then use low-dimensional encoders for text and to calculate similarities. A generator module can be added in retrieval augmented generation (RAG) with a generative language model to align the outputs with the input question in a final step \cite{lewisRetrievalAugmentedGenerationKnowledgeIntensive2021}. Despite the increasing number of parameters, generators are not standalone models as they lack ground knowledge \cite{maLargeLanguageModel2023}\cite{sunHeadtoTailHowKnowledgeable} and tend to fabricate information (``hallucination''). Recent language models can manipulate both natural language and programming languages \cite{touvronLlamaOpenFoundation2023}, which increases the capacities of information retrieval systems.


\paragraph*{An example of machine learning for street names.} Despite its limited information in comparison with DBpedia, Wikidata is an atomic knowledge graph used in StreetToPerson \cite{gurtovoyLinkingStreetsOpenStreetMap2022}. This model restricts the candidates for the origin of place names to persons only: given a place name \textit{X Street}, StreetToPerson extracts vectors of attributes (e.g., name, occupation, place) for persons having \textit{X} in their name and then trains a random forest classifier.

\section{Methodology and architecture}
\label{sec:method}

Our model is designed to identify the origin of a place name by extracting related knowledge from a spatial knowledge graph.

\paragraph*{Anchor-question.}  An example question that our model aims to solve is: ``Who is the person that Batman Street in Melbourne is named after?'' We formulate such a question into the following format:

\begin{quote}
    \textit{``\colorbox{Apricot!40}{Who}$^{(A)}$ is $[X_{place}]$ most likely \colorbox{Apricot!40}{named after}$^{(A)}$, \colorbox{LimeGreen!40}{in $[city([X_{place}])]$}$^{(B)}$, \colorbox{SeaGreen!40}{$[country([X_{place}])]$}$^{(C)}$? If it is not a \colorbox{Apricot!40}{person}$^{(A)}$, find any other \colorbox{Apricot!40}{origin}$^{(A)}$. "}
\end{quote}

Here, $X_{\textrm{place}}$ represents the target place name. As our model is based on language model generation, the template above will focus more on a person origin of $X_{\textrm{place}}$ (\colorbox{Apricot!40}{A}). It also implies a spatial context involving a coarse spatial filter (\colorbox{SeaGreen!40}{C}) and a finer-grained spatial filter (\colorbox{LimeGreen!40}{B}).

To retrieve the answer, the model first uses a \textbf{searcher} to extract candidate answers from DBpedia. Then, it uses an \textbf{indexer-ranker} to rank these candidates based on their closeness with the query. Finally, a \textbf{generator} module produces the final answer from the ranked list of candidates. These three modules are detailed below, with reference to the overall architecture illustrated in Figure~\ref{fig:general_schema}. 
 
\subsection{Searcher.}
Given a query for the origin of a place name, the searcher module aims to extract any relevant data from an external spatial knowledge graph. In this work, we used DBpedia as the external knowledge source, where all knowledge is stored as triplets, i.e.,  (\textit{subject, predicate, object}). Let $X_{\textrm{place}}$ be the place name mentioned in the query. We used the DBpedia SPARQL endpoint to extract $k_{\textrm{searcher}}$ triplets from the DBpedia knowledge graph. Each extracted triplet satisfies two constraints: (1) the subject must contain the place name $X_{\textrm{place\_name}}^*$; and (2) the predicate must belong to a predefined relation set $F_{\textrm{rel}}$, where $F_{\textrm{rel}} = \{$\textit{abstract, children, comment, country, date, geo, label, location, occupation, parent, place, spouse}$\}$. In particular, \{\textit{abstract}, \textit{comment}\} are included in $F_{\textrm{rel}}$ because triplets with these predicates usually include rich descriptions about their subjects. Extracted triplets are stored in a RDF/XML format. In order to improve the compactness of RDF/XML, all the URIs\footnote{Uniform Resource Identifier} were mapped with their prefixes, allowing LMs used in later modules to focus on those meaningful tokens in the URIs.

\subsection{Indexer and ranker.}
An indexer and a ranker were developed to identify those triplets that are more relevant to the query. We developed the indexer and ranker based on ColBERTv2 \cite{santhanamColBERTv2EffectiveEfficient2022}, a pre-trained LM designed to compute the closeness between a pair of query and a triplet document in terms of semantic similarity. Here, the texts associated with a set of triplets related to one subject can be treated as one triplet document, and thus the semantic similarity between a query and a triplet document is measured as the similarity between their latent embeddings produced by a text encoder. ColBERTv2 enhances the efficiency of similarity computation by grouping and indexing the triplet documents into several clusters, where passages within the same cluster are more similar to each other. The ranker ranks all triplet documents based on their similarities to the query and returns the top-$k_{\textrm{ranker}}$ documents as the result. The model is trained to rank documents related to the query (positive samples) higher than the documents that are irrelevant to the query (negative samples).

We chose ColBERTv2 because it offers a balance between language understanding and computational efficiency. However, most LMs are not specifically trained to understand spatial concepts; ColBERTv2 is trained on general knowledge such as texts crawled from Wikipedia. To answer spatial queries, a system needs to handle spatial concepts accurately. Continuing our example query, ``Who is the person that the Batman Street in Melbourne is named after?''; an accurate query response should respect common spatial knowledge such as ``Melbourne is contained in Victoria'' and ``Melbourne is a city in Australia.'' To inject spatial knowledge to the indexer-ranker, we fine-tuned the base model ColBERTv2 on two datasets curated for spatial understanding: \DSgn and \DSgncountry. 

Further, we note that the base ColBERTv2 model was pre-trained on pure natural language text. Although RDF/XML format presents a structure that is mostly human readable, it still involves RDF/XML-specific syntax in its format, which may be harder for the base model to interpret. Hence, we fine-tuned the indexer-ranker on a dataset that we curated, namely \DSqald.

\subsubsection*{Fine-tuning the indexer-ranker.}
We fine-tuned the base language model ColBERTv2 using three datasets. \DSgn and \DSgncountry were utilized for improving spatial understanding, while \DSqald was used for improving the understanding of RDF/XML formats.

\paragraph*{\textbf{\DSgncountry}.} This dataset contains questions and answer pairs about countries that are neighboring to one another. We extracted countries from Geonames and constructed a graph $G^{\textrm{country}} = \{V^{\textrm{country}}, E^{\textrm{country}}\}$. Here, $V^{\textrm{country}}$ is the node set with each node representing one country, and $E^{\textrm{country}}$ represents the set of edges. Given country $[\textrm{country}_i]$ and $[\textrm{country}_j]$ that share borders, we added an edge $e_{i,j}^{\textrm{country}}$ to set $E$ to represent the adjacency relationship. We generated one question-answer pair for edge $e_{i,j}^{\textrm{country}}$ using the following template:

\begin{equation}
    \left\{
    \left(
    \begin{array}{ccl}
        X & = & \textrm{`` Give a country that shares a border with $[\textrm{country}_i]$. ''}\\
        y^+ & = & \textrm{`` $[\textrm{country}_j]$ shares a border with $[\textrm{country}_i]$."}  \\ 
    \end{array}
    \right)
    \right\}
    _{\forall e_{i,j}^{\textrm{country}} \in  E^{\textrm{country}}}
    \label{eq:1}
\end{equation}
Here, $y$ represents a country that is a \textit{positive} answer to the question $X$. To fine-tune ColBERTv2, we also generated negative samples to the same question $X$, so that the model can learn to discriminate the spatial context involved in $X$. Specifically, we randomly sampled a country $[\textrm{country}_k]$ that does not share border with $[\textrm{country}_i]$ and generated a negative sample:
\begin{equation}
    \begin{array}{ccl}
        y^- & = & \textrm{`` $[\textrm{country}_k]$ shares a border with $[\textrm{country}_i]$."}  \\ 
    \end{array}
\end{equation}

\paragraph*{\textbf{\DSgn}.} This dataset completes \DSgncountry by also capturing the closeness among cities. We extracted cities from Geonames, keeping only cities with a population of at least $n_{\textrm{hab}}$. For these cities, we computed their pairwise spherical distance. Similar to \textbf{\DSgncountry}, we constructed a city graph denoted as $G^{\textrm{city}} = \{V^{\textrm{city}}, E^{\textrm{city}}\}$ where each node represents one city. Given city $[\textrm{city}_i]$ and $[\textrm{city}_j]$, if their distance is less than or equal to the threshold $d_{\textrm{city}}$, we added an edge $e_{i,j}^{\textrm{city}}$ to the edge set, and generated a question answer pair using the following template:  
\begin{equation}
        \left\{
        \left(
        \begin{array}{ccl}
            X & = & \textrm{``Give a city near $[\textrm{city}_i]$ in $[\textrm{country}(\textrm{city}_i)]$."}\\
            y^+ & = & \textrm{``$[\textrm{city}_j]$ in $[\textrm{country}(\textrm{city}_j)]$ is a neighbor of $[\textrm{city}_i]$ in $[\textrm{country}(\textrm{city}_i)]$."}  \\ 
        \end{array}
        \right)
        \right\}
        _{(i,j)}
    \label{eq:2}  
\end{equation}
The negative samples in \DSgn were generated in a similar way to \DSgncountry. We omit the details for conciseness.

\paragraph*{\textbf{\DSqald}.} This dataset covers questions on general knowledge rather than focusing on spatial knowledge. Given a question $X$, its answer is curated by extracting sub-knowledge graphs from DBpedia in RDF/XML format. We built this dataset based on QALD9, which includes a range of questions and their corresponding SPARQL queries on DBpedia \cite{usbeck9thChallengeQuestion2018}, denoted as $\{(X, y_{\textrm{SPARQL}})\}$. Given the $i$-th pair consisting question $X_i$ and query $y_{\textrm{SPARQL}, i}$, we executed the query and extracted the knowledge graph $y_{\textrm{KG},i}^{+}$, forming the new pair $\{(X_i, y_{\textrm{KG}, i}^{+}) \}$. For training, we also identified negative samples $y_{\textrm{KG},i}^-$ by retrieving nodes in DBpedia that contained one keyword (provided by QALD9) in the query, but that are not returned by executing $y_{\textrm{SPARQL}}$.

\subsection{Generator}
From the top-$k_{\textrm{ranker}}$ documents, a generative language model, Llama2 \cite{touvronLlamaOpenFoundation2023}, was used to choose the top-1 document and generate the final answer.

The top-$k_{\textrm{ranker}}$ documents were concatenated in the prompt. To concatenate these documents, we experimented with two ways of ordering these documents: ordering by increasing similarity with the query \sortDownTop and decreasing rank \sortTopDown. In our experiments, we observe that positioning the most relevant information near the tail of the prompt improves the results: due to limitations with long-distance dependencies, the model focuses on the most recent input to the language model, i.e., the tail of the prompt. We applied in-context learning into the design of our prompt, where an example of how to solve the task is provided to the language model. The prompt is designed as follows to retrieve the top-$k_{\textrm{generator}} (k_{\textrm{generator}} = 1)$ document. 

\begin{tcolorbox} [width=\linewidth, sharp corners=all, colback=white!95!black]
\small
For [$k_{\textrm{ranker}}$] knowledge graphs, an extract from the RDF/XML file is provided; the names of the knowledge graphs are: [subject$_1$]$\ldots$[subject$_{k_{\textrm{ranker}}}$]. We want to find an answer to: ``[ANCHOR QUESTION]". These are the extracts: [KG$_1$]$\ldots$[KG$_{k_{\textrm{ranker}}}$]". Give me a simple answer to ``[ANCHOR QUESTION]". [INSTRUCTIONS\footnote{Instructions: ``Only use the provided information. First, choose the extract that best allows you to answer among: [title$_1$]$\ldots$[title$_{k_{\textrm{ranker}}}$]. Delimit your chosen answer with the tags $\langle$CHOICE$\rangle$  $\langle$/CHOICE$\rangle$ . Second, give your answer by delimiting it with the tags $\langle$ANSWER$\rangle$  $\langle$/ANSWER$\rangle$. Your answer should be concise. If it is a person, I need the first name and the last name. For example, to ``Who is Rue Madame Curie in Beirut, Lebanon named after?", write: ``$\langle$CHOICE$\rangle$  [write\_here\_your\_chosen\_source] $\langle$/CHOICE$\rangle$  $\langle$ANSWER$\rangle$  Marie Curie $\langle$/ANSWER$\rangle$  Based on the provided information, ..."}]
\end{tcolorbox}

Here, the generated prompts were sent to a non fine-tuned Llama2-13B-Chat model, frozen with a 4-bit quantization. This maximum size of context is 4096, allowing us to concatenate all the chosen documents without truncation.

\section{Experimental design}

Each place name $X_{\textrm{place}}$ in the gazetteer is treated independently as a sample: the results of one sample is not re-used for another place name.

\subsection{Datasets}
We curated a dataset for evaluating the presented framework. Let \textbf{\DSgaz} $= \{(X_{\textrm{place}}^i, y_{\textrm{origin}}^i)\}_{i=1}^N$ be the dataset containing $N$ data samples, where the $i$-th sample is a pair of location $X_{\textrm{place}}^i$ and its ground-truth origin $y_{\textrm{origin}}^i$. This dataset is derived from the \textit{Vicnames}, a comprehensive database containing rich information about place names owned by Victorian State Government register of Geographic Place Names, Australia. In this work, we filtered this dataset to keep only street names in the city of Melbourne to perform a focused study. In addition, for each place name, we extracted its root name by removing any prefix and street type, e.g., converting \textit{Little Bourke Street} to \textit{Bourke}. The resultant dataset contains 248 entries, i.e., $N=248$.

As mentioned in Section~\ref{sec:method}, three datasets are curated to fine-tune the underlying language model, ColBERTv2, of the indexer-ranker: \DSgncountry, \DSgn, and \DSqald. To prepare \DSgncountry, we kept the ratio between positive and negative answers as 1:100. To prepare the dataset \DSgn, we set $n_{\textrm{hab}} = 50\textrm{K}$, $d_{\textrm{city}} = 50$~km. For each question, we kept the ratio between positive and negative answers as 1:5 at maximum. The statistics of these datasets are presented in Table \ref{tab:datasets_count}.

\subsection{Evaluation}
Using \DSgaz, we evaluate the model in terms of its capability in two aspects: (1) understanding semantic meaning of a query; and (2) processing spatial contexts involved in a query. The following measures are utilized to evaluate the relevance of the retrieved answer to the query: 
\begin{itemize}
    \item \textbf{Semantic relevance: \textit{sem}.} In this measurement, we focus on the semantic similarity of the retrieved answer and the textual description of the place name $X_{\textrm{place}}$ in the gazetteer. Suppose that $X_{\textrm{place}}$ is \textit{Nancy Adams Place} in Melbourne that is named after a local person in Melbourne. We evaluate if the retrieved answer is semantically relevant to $X_{\textrm{place}}$. For example, if the system identifies the origin as \textit{Nancy Adams} who lived in Victoria, then the answer is semantically overlapped with the query, and hence, it is deemed as semantically correct to the query. In contrast, the botanist \textit{Nancy Adams} in New Zealand is considered as erroneous. 
    \item \textbf{Spatial relevance: \textit{geo\_aus}, \textit{geo\_vic}.} Similarly, we measure the spatial similarity of the retrieved answer and the place name $X_{\textrm{place}}$ in the query. Specifically, if the retrieved origin is related to the spatial context of $X_{\textrm{place}}$ (e.g., both retrieved origin and $X_{\textrm{place}}$ are related if they share a same spatial parent from a hierarchical spatial representation for containment; the parent can be the ``name of the country"~-- equal to Australia for \textit{geo\_aus}~-- or the ``name of the state"~-- equal to Victoria for \textit{geo\_vic}), we treat it as a correct answer. 
\end{itemize}

The evaluation is conducted by one annotator.
We report the accuracy of each module in our framework in terms of HR$@k$: for $N$ data samples in \textbf{\DSgaz}, on average, what is the probability of the ground-truth answer being in the top-$k$ retrieved answers. Since the indexer-ranker provides an ordered list of retrieved answers, we report the quality of the returned list by reporting three more metrics: mean reciprocal rank ($\textrm{MRR}@k$), normalized discounted cumulative gain ($\textrm{nDCG}@k$), and precision$@k$. Here, the $\textrm{MRR}@k$ computes the average rank of the ground-truth item within the list of top-$k$ answers; the $\textrm{nDCG}@k$ is similar to hit ratio but penalizes the result if the ground-truth answer is ranked low; the precision$@k$ ($\textrm{P}@k$) reports the average number of answers that are related to the query.

As benchmark models, the results after the generator are compared to the scores given by gpt-4o-mini and StreetToPerson \cite{gurtovoyLinkingStreetsOpenStreetMap2022}. 

\begin{table}
    \centering
    \caption{Statistics of the three datasets used for fine-tuning the indexer-ranker. In GeoNames, overseas territories of a country are considered as an independent entry for countries (e.g. French Guiana and France have two distinct entries), which explains the high number of countries.}
    \begin{tabular}{llllll}
        \toprule
        & Dataset & Positive pairs & Negative pairs & Comments\\
        \midrule
        \DSqald & \faComments & 647  & 64 401 & 408 questions\\
        \DSgncountry & \faGlobeAsia & 650 & 25 751 & 252 "countries"\\
        \DSgn & \faGlobeAsia \hspace{1ex} \faCity & 320 958 & 746 938 & 252 "countries" + 10~572 cities\\
        \bottomrule
    \end{tabular}
    \label{tab:datasets_count}
\end{table}

\subsection{Experiments}

\paragraph*{Searcher.} We only extracted triplets whose language of the object is English or not specified. The Virtuoso SPARQL endpoint for DBpedia was used with a limitation of $k_{\textrm{searcher}} =$ 10K triplets. Additionally, we fixed a maximum of 1000 subjects. For reproducibility, the query was executed once, with all extracted knowledge graphs saved locally for all the experiments. 

\paragraph*{Indexer and ranker.} We set $k_{\textrm{ranker}}$ as 10, meaning that the indexer-ranker will select the top-$10$ triplets from the 10K triplets retrieved from the searcher. The knowledge graph extracted by the searcher was split by subject in the triple; each chunk is regarded as a \textit{document} for ColBERTv2. The maximum length for input is set to 256 tokens.
As presented in Table \ref{tab:models_colbert}, we developed five versions of ColBERT (C0, C1, C2a, C2b, C2c) that underwent different fine-tuning procedures. C0 [\symbCZero] is the original ColBERT without fine-tuning; C1 [\symbCOne] is fine-tuned on \DSqald; C2a [\symbCTwoA] is fine-tuned on \DSgncountry and then \DSqald; C2b [\symbCTwoB] is fine-tuned on \DSgn and then \DSqald; and C2c [\symbCTwoC] is fine-tuned on both \DSgn and \DSqald, where the two datasets are fed into ColBERTv2 simultaneously with shuffling.

\begin{table}[H]
    \centering
    \caption{Versions of ColBERT that are fine-tuned on different combination of datasets, where $\varnothing$ represents the corresponding dataset is not utilized. The notations \textcircled{1} and \textcircled{2} represent that the corresponding dataset is utilized for fine-tuning, where 1/2 represents the order of the dataset introduced during fine-tuning.\label{tab:models_colbert}}
    \makebox[\textwidth]{
    \begin{tabular}{cc|ccc}
        \multicolumn{2}{c}{\textbf{Version of}} & \multicolumn{3}{c}{\textbf{The fine-tuning is carried out on the following datasets:}} \\
        \multicolumn{2}{c}{\textbf{ColBERT}} & \DSgncountry \faGlobeAsia & \DSgn \faGlobeAsia \hspace{1ex} \faCity & \DSqald \faComments \\
         \toprule
        C0 & {\tiny [\symbCZero]} & $\varnothing$ &  $\varnothing$ &  $\varnothing$ \\
        C1 & {\tiny [\symbCOne]} & $\varnothing$ &  $\varnothing$ &  \textcircled{1} \\
        C2a & {\tiny [\symbCTwoA]} & \textcircled{1} &  $\varnothing$ &  \textcircled{2} \\
        C2b & {\tiny [\symbCTwoB]} & $\varnothing$ &   \textcircled{1} &   \textcircled{2}\\
        C2c & {\tiny [\symbCTwoC]} & \textcircled{1} &  $\varnothing$ &   \textcircled{1}\\
    \end{tabular}
    }
\end{table}

\paragraph*{StreetToPerson.} This model successfully extracted 261 compatible pairs ``street-person'' in Australia from the English Wikidata and Wikipedia as a training dataset. Even if the streets of Melbourne are part of the test dataset, we decided to include the 22 streets detected in Melbourne in the training dataset for three reasons. First, we want to characterize the streets in Melbourne to improve inference, hence it is necessary to have data for this area. Second, the dataset for Australia is 18 times smaller than in the original paper for Germany. Third, our objective is not to find origins \textit{ex nihilo} but rather to detect more origins than those already recorded in gazetteers: we accept that the model is biased by having seen the training dataset. Moreover, the pre-trained LMs used in our model are also biased as they might have seen also the training dataset. The parameters for StreetToPerson are unchanged. 



\section{Results}

\begin{table}
    \centering
    \caption{Description of \DSgaz}
    \begin{tabular}{p{1ex}p{1ex}p{5cm}|rr}
        \toprule
         &  &  & \textbf{Count} & \textbf{\%} \\
        \midrule
        \multicolumn{3}{l|}{Number of streets} & 248 & \\
         & \multicolumn{2}{l|}{... with an origin in the gazetteer} & 230 / 248 & .927 \\
         & \multicolumn{2}{l|}{... that commemorates a named person} & 143 / 248 & .577 \\
         & \multicolumn{2}{l|}{... that commemorates an unnamed person} & 68 / 248 & .274\\
         & \multicolumn{2}{l|}{... where the searcher successfully extracts a knowledge graph} & 222 / 248 & .895 \\
         &  & ... and whose origin is mentioned in the knowledge graph ($\textrm{HR}@10\textrm{K}$ for the searcher) & 93 / 222 & .375 \\
         \bottomrule
    \end{tabular}
    \label{tab:desc_ds_gaz}
\end{table}

\subsection{Searcher}

Table \ref{tab:desc_ds_gaz} presents the results for the searcher. The gazetteer provides 248 streets in \DSgaz with a known origin in 92.7\% of the cases. Only 57.7\% of the streets commemorate a person. Of those, 27.4\% commemorate a local inhabitant of the area, such as a merchant or a former land owner, even though they are not explicitly named in the document\footnote{For example, where the documentation specifies ``this street is named after a former person who lived in the street'' but does not explicitly name that person. This qualification was subjectively defined by the annotator.}. The searcher successfully extracts a knowledge graph from DBpedia for 89.5\% of the streets: the missing graphs are due to a lexical gap (e.g., \textit{Abeckett} Street in the gazetteer instead of \textit{A'Beckett} Street) or the absence of information in DBpedia. On average, the resulting dataset has 291 objects per knowledge graph. 

Among the extracted knowledge graphs, only 37.5\% ($\textrm{HR}@10k$) contain a mention of the origin (see Table \ref{tab:res_gir_models}). A \textit{mention} does not necessarily mean that there is an explicit link between a candidate and the naming origin; only that the origin appears in the text. This relatively low score for the searcher indicates that not all the required information is easily accessible in DBpedia, first due to the limitations of the SPARQL endpoint and second due to the prevalence of unnamed persons.

\subsection{Ranker}

\begin{figure}
\centering

    \makebox[\textwidth]{
        \subfigure[C0~(\symbCZero) - not trained on a spatial dataset]{
            \includegraphics[width=\textwidth]{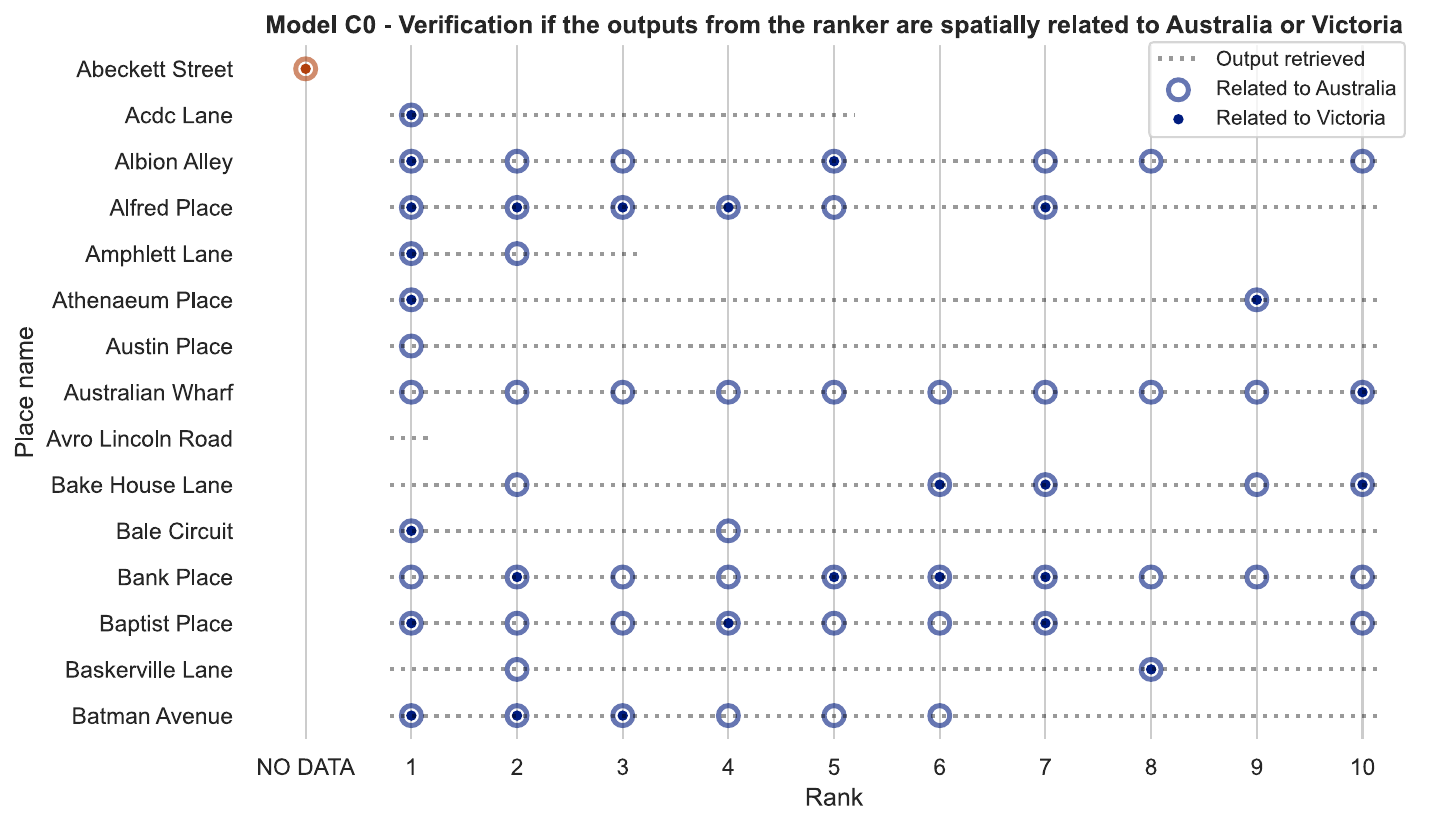} 
            \label{fig:distrib_C0}
        }
    }

    \makebox[\textwidth]{
        \subfigure[C2a~(\symbCTwoA) - trained on a spatial dataset]{
            \includegraphics[width=\textwidth]{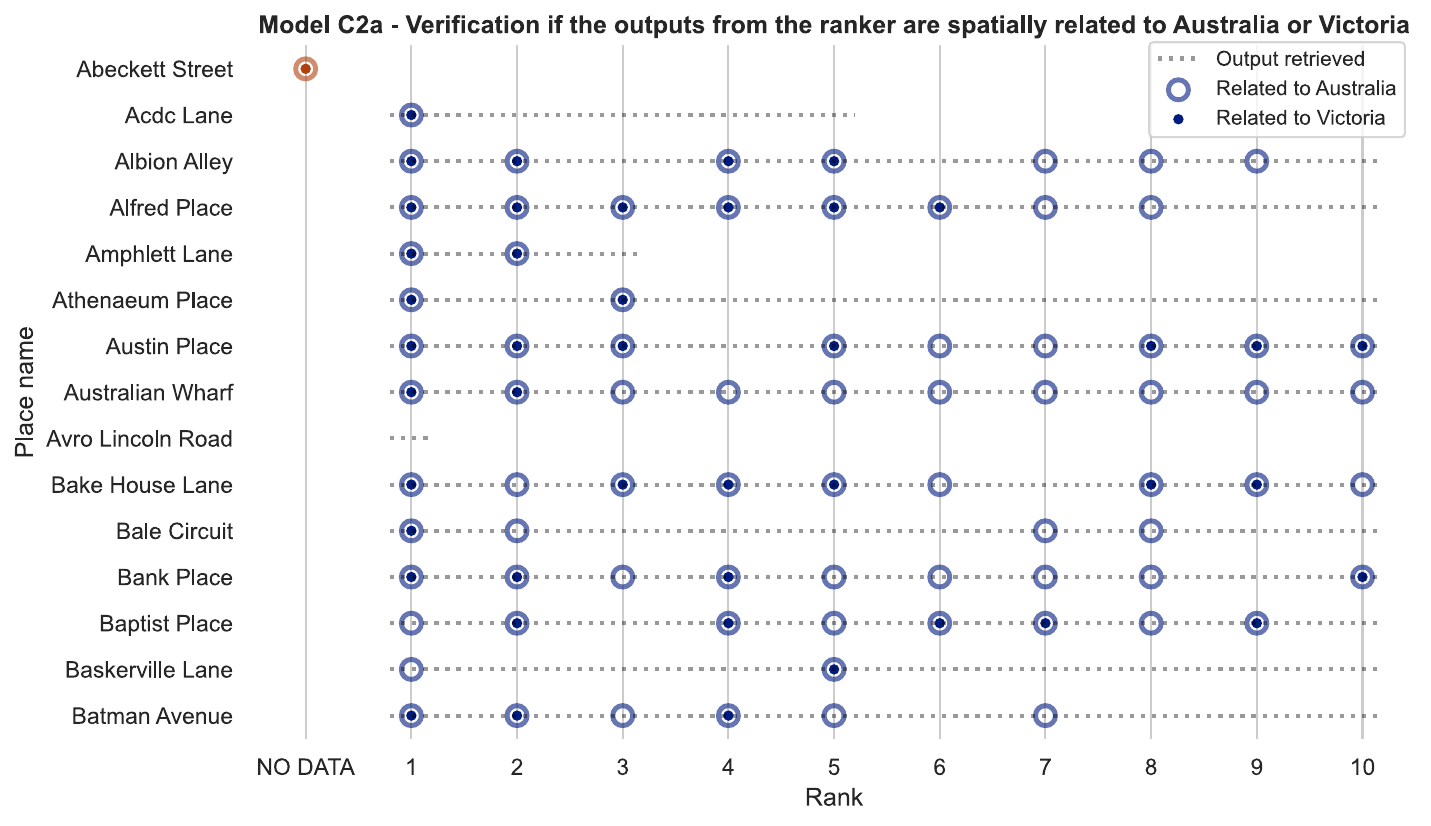}
            \label{fig:distrib_C2a}
        }
        
    }

\caption{Retrieved items for 15 place names regarding \textit{geo\_aus} and \textit{geo\_vic}. A fine-tuning is expected to highlight spatially related candidates at the highest ranks, which is characterized by the $\textrm{nDCG}@10$.}
\label{fig:distrib_all}
\end{figure}

Table \ref{tab:res_gir_models}  and Figure~\ref{fig:distrib_all} present the results for the indexer-ranker, where the subscript $sem$ represents evaluation results in terms of semantic understanding and the subscript $geo\_vic$ and $geo\_aus$ represent results in terms of spatial understanding (i.e., if the answer is respectively 
at least within Victoria or Australia). 


\newcommand{\descTableCzero}{\multirow{3}{*}{\makecell{C0 \\ {\tiny [$\varnothing$]}}}}
\newcommand{\descTableCone}{\multirow{3}{*}{\makecell{C1 \\ {\tiny [\faComments]}}}}
\newcommand{\descTableCtwoA}{\multirow{3}{*}{\makecell{C2a \\ {\tiny $[$\faGlobeAsia $]$} \\ {\tiny $+ [$ \faComments $]$} }}}
\newcommand{\descTableCtwoB}{\multirow{3}{*}{\makecell{C2b \\ {\tiny $[$ \faGlobeAsia \hspace*{.1ex} \faCity $]$}  \\ {\tiny $+$ \hspace*{.1ex}$[$\faComments $]$}}}}
\newcommand{\descTableCtwoC}{\multirow{3}{*}{\makecell{C2c \\ {\tiny $[$ \faGlobeAsia \hspace*{.1ex} \faComment $]$} }}}
\newcommand{\midsep}{\cdashline{1-10}[.5pt/2pt]}
\newcommand{\midsepbis}{\cdashline{1-4}[.5pt/2pt]}

\begin{table}
    \centering
    \caption{Evaluation of the performances of C0, C1, C2a, C2b and C2c after the ranker ($@k = @10$) for the different types of observations (relevance of the semantic evaluation \textit{sem}, or spatial evaluations \textit{geo\_aus} or \textit{geo\_vic}). First, the scores are calculated for all the data in \DSgaz and second, on the a subset where the extracted knowledge graph KG from the searcher contains a mention of the origin: these scores are marked with $^{*}$. The notation $\overline{x}$ denotes averaged results on the dataset; MAP is the mean average precision (namely $\overline{P}@10$). Each street in \DSgaz ($N=248$) represents one independent sample; results are averaged on $N=248$ elements.}
        \begin{tabular}{cl|llll|llll}

            \toprule
            \multicolumn{2}{c|}{Ranker} & \multicolumn{4}{c|}{\textbf{Scores on \DSgaz ($N=248$)}} & \multicolumn{4}{c}{\textbf{... restricted to the streets where}}\ \\
            \multicolumn{2}{c|}{$@k = @10$} & \multicolumn{4}{c|}{} & \multicolumn{4}{c}{\textbf{the KG mentions an origin ($N^{*}=93$)}}\ \\
            Model & Type & $\overline{\bm{\textrm{MRR}}}$ & $\overline{\bm{\textrm{nDCG}}}$ & \textrm{MAP} & \textrm{HR} & $\overline{\bm{\textrm{MRR}}}^{*}$  & $\overline{\bm{\textrm{nDCG}}}^{*}$  & \textrm{MAP}$^{*}$ & \textrm{HR}$^{*}$ \\

            \midrule
            
            \descTableCzero & sem & .232 & .170 & .063  & .216 & .445 & .412 & .143 & .506 \\
            & geo\_aus & .684 & .777 & .501 & .883 & .831 & .854 & .578 & .933 \\
            & geo\_vic & .476 & .539 & .243 & .703 & .628 & .646 & .313 & .753 \\
            \midsep

            
            \descTableCone & sem & \best{.253} & \best{.186}  & \best{.080} & \best{.243} & \best{.475} & \best{.421} & \best{.177} & \best{.528} \\
            & geo\_aus & .769 & .854 & .588 & .896 & .885 & .908  & .656  & .944 \\
            & geo\_vic & .585 & .637 & .294 & .739 & .728 & .729  & .356  & .786 \\
            \midsep

            
            \descTableCtwoA & sem & .229 & .162 & .076 & .221 & .431 & .382  & .173 & .506 \\
            & geo\_aus & .780 & .858 & .585 & .892 & .889 & .909 & .664 & .944 \\
            & geo\_vic & .602 & .657 & .306 & \best{.748} & \best{.740} & \best{.745} & .366 & \best{.798} \\
            \midsep

            
            \descTableCtwoB & sem & .237 & .169 & .071 & .213 & .450 & .400  & .160 & .489 \\
            & geo\_aus & \best{.797} & \best{.891} & \best{.635} & .900 & .887 & \best{.934} & \best{.699} & \best{.955} \\
            & geo\_vic & .555 & .620 & .283 & .733 & .673 & .698 & .340 & .773 \\
            \midsep

           
            \descTableCtwoC & sem & .232 & .166 & .073 & .225 & .418 & .366 & .159 & .483 \\
            & geo\_aus & .792 & .881 & .618  & \best{.901} & \best{.890} & .926 & .679 & \best{.955} \\
            & geo\_vic & \best{.615} & \best{.674} & \best{.318} & .743 & .717 & .743  & \best{.375} & .786 \\

            \bottomrule
        \end{tabular}
    \vspace{1ex}

    \label{tab:res_gir_models}

\end{table}

\paragraph*{General.}
Initially pre-trained on English, a non fine-tuned ColBERTv2 C0 [\symbCZero] can already understand the RDF/XML format of knowledge graphs. However, the quality of the top-$k$ candidates is not high with respect to their spatial distribution, particularly with the $\overline{\textrm{nDCG}}$ compared with fine-tuned models. To evaluate the models in detail, we must distinguish two cases: firstly, can a model retrieve an information that does exist? Secondly, if missing, can a model retrieve information that is at least spatially related to the query? 

\paragraph*{When a knowledge graph mentions an origin.}
In these cases, evident answers are already strong markers, without the need to attend to spatial understanding. 

Indeed, the models does not discard the answer for the query in the top-10 in more than 48\% of the cases, as indicated by the $\textrm{HR}_{\textrm{sem}}^{*}@k$. Moreover, $\overline{\textrm{MRR}}_{\textrm{sem}}^{*}$ is over 41\% which means that the first evident answer appears quickly in the highest ranks. We observe that $\textrm{MAP}_{\textrm{sem}}^{*}$ is between .1 and .2: an interpretation is that, on average, 1 or 2 retrieved items among the set of top-10 contain relevant information. Nonetheless, certain names like \textit{Flinders} Street are more specific than \textit{Rainbow} Alley, which increases the chances to have more than 2 items instead of 0. Mentions to \textit{Australia} or \textit{Victoria} are good markers as the top-10 is likely to contain at least one related element in more than 75\% of the recommendations when we consider HR$_{\textrm{geo\_vic}}^{*}$ or 93\% with HR$_{\textrm{geo\_aus}}^{*}$, as well as the first two elements out of 10 might be linked to these places ($\overline{\textrm{MRR}}_{\textrm{geo\_aus}}^{*}@10 \geq .6$).

\paragraph*{When the knowledge graph does not mention any relevant origin.} 
In these cases, the \textit{sem} score has no useful information but in contrast, the ranker is expected to manipulate more information related to \textit{geo\_aus} or \textit{geo\_vic}. In practice, we observe that the models do not fully exploit the spatial filters written in the context (in the anchor-question) as a discriminating criterion. 

An efficient ranker should first prioritize passages that do contain the origin of a place name. This origin can be explicitly given in the passage, or inferred with multiple passages eventually with the internal knowledge of language models. As a second order of priority, human understanding would focus on spatial similarities, as indicated in the anchor-question, by giving better ranks for items related to Melbourne, Victoria, or Australia. This characteristic is not fully respected by our results. For all the models, the scores for the $\overline{\textrm{nDCG}}_{\textrm{geo\_...}}@k$ and $\textrm{MAP}_{\textrm{geo\_...}}@k$ are worse than $\overline{\textrm{nDCG}}_{\textrm{geo\_...}}@^{*}k$ and $\textrm{MAP}_{\textrm{geo\_...}}^{*}$ with a systematic difference of more than .05. This indicates that the models tend to provide more attention to the semantic aspect than the two spatial in-context filters. In other words, mentioning \textit{in Melbourne, Victoria, Australia} in the anchor-question does not act as a filter.


\paragraph{Effect of training.} 
In these cases, fine-tuning is mainly useful to read the RDF format and to find similarities based of the global meanings, with a compromise with spatial understanding.

Training with \DSqald offers better models for the evaluation \textit{sem} than the baseline C0~[\symbCZero] since ColBERT now understands the RDF/XML format. However, there is a compromise between the semantic and the spatial scores: a previous fine-tuning on \DSgncountry and \DSgn reduced the scores on \textit{sem} to offer better scores for \textit{geo\_aus} or \textit{geo\_vic} as shown in Figure~\ref{fig:distrib_all} for example. As a surprising result, training on a fine-grained grid of locations with \DSgn does not improve scores on fine locations \textit{geo\_vic} but only on coarse locations with \textit{geo\_aus}. We propose two explanations to this observation: first, \DSgn contains more mentions to countries that strengthen the similarities between countries and second, the mention of one city is long-tail information that has little impact on the back propagation of the loss. In contrast, only training on \DSgncountry at a coarser spatial level keeps a certain capacity for the language models to generalize. In Figure~\ref{fig:distrib_all}, we show that the fine-tuning on spatial pairs does not fully improve the rankings: qualitatively, the distribution of the results still keeps a high entropy. We then assume that a continual learning first on spatial then RDF/XML understanding may lead to ``catastrophic forgetting'' of spatial knowledge in C2a~[\symbCTwoA~] and C2b~[\symbCTwoB~]: that is why we define C2c~[\symbCTwoC~] with a unique training on both skills. Finally, C2c~[\symbCTwoC~] does not necessarily lead to better performance. By consequence, the paradigm of fine-tuning hardly captures both semantic and spatial understanding in a neural information retrieval only based on a language model.



\subsection{Generator}

\begin{figure}[htb]
\centering

        \subfigure[Scores on \DSgaz ($n=248$)]{
            \includegraphics[scale=.75]{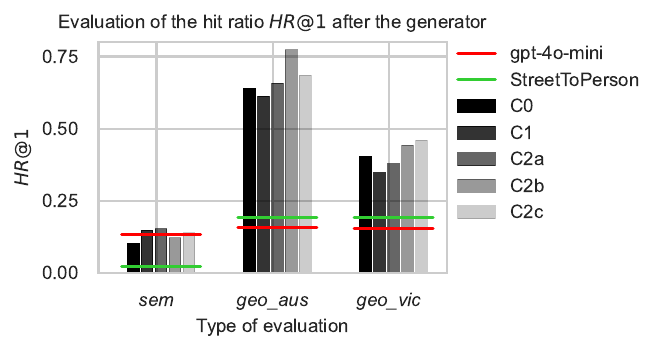} 
            \label{fig:hr1}
        }  


        \subfigure[... restricted to the streets where the KG mentions an origin ($n^{*}=93$)]{
            \includegraphics[scale=.75]{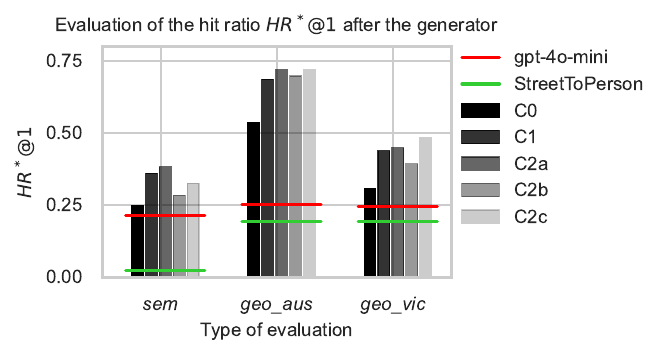} 
            \label{fig:hr1Restrict}
        }   

        

\caption{Evaluation of the hit ratio on the types \textit{sem}, \textit{geo\_aus} and \textit{geo\_vic} after the generator for the models C0, C1, C2a, C2b, C2c and comparison with the baselines gpt-4o-mini and StreetToPerson.}
\label{fig:resGenerator}
\end{figure}

\begin{figure}[htb]
\centering
      \makebox[\textwidth]{\includegraphics[scale=.75]{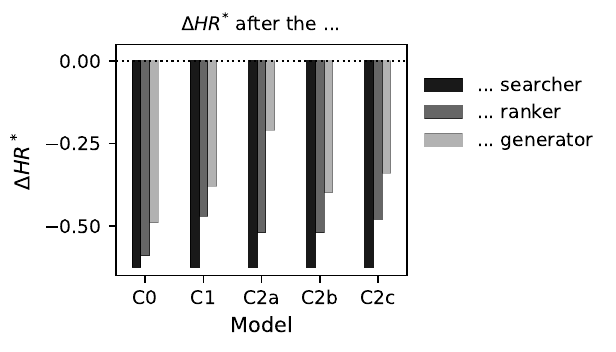}}
    \vspace{-5ex}
\caption{Relative change $\Delta \textrm{HR}^*(\textrm{module}_i, \textrm{module}_{i-1})$ with $\textrm{module}_i \in \{$ searcher, ranker, generator $\}$. The more $\Delta \textrm{HR} \rightarrow 0$, the more relevant information is selected without loss.}
\label{fig:loss_modules_IR_graph}
\end{figure}

Figure~\ref{fig:resGenerator} and Figure~\ref{fig:loss_modules_IR_graph} present the results for the generator, discussed further below.

\paragraph*{Comparison with the baselines.} StreetToPerson does not generalize to Australia; gpt-4o-mini provides insightful answers only for common knowledge whereas our models offer better and specialized answers in the long-tail knowledge. Indeed, for StreetToPerson, $\textrm{HR}_{...}@1$ and $\textrm{HR}_{...}^*@1$ are significantly lower than in our approach. With gpt-4o-mini, $\textrm{HR}_{\textrm{sem}}@1$ is better than StreetToPerson but still lower than our approach; moreover, gpt-4o-mini correctly retrieves an origin for streets that are named after common elements (Plover\footnote{A bird} Lane or Bridleway\footnote{Path originally used by people riding horses or trails} Walk) contrary to our model.

\paragraph*{Capacity of generalization.} With limited supervision, the use of a generator aims to align a final answer with an initial question, while discriminating information. In our experiments, the generator fills this role but is not optimized to take decisions. 

The generator tends to systematically answer the query without hallucination; however, it seldom rejects an irrelevant top-$k_{\textrm{ranker}}$ set: we only encounter the answer ``There is no relevant information to answer the question'' once or twice. In our experiments, limited scores for the $\textrm{HR}@1$ in  Figure~\ref{fig:resGenerator} are found. We believe the reasons can be twofold. First, the model does not understand the background knowledge of naming conventions, e.g., some rules about \textit{how} places are named. During the evaluation, the model tends to over-estimate persons from the world of sport, particularly football and rugby. Second, the information ground-truth answer may be absent in the dataset and the model has to infer from the existing knowledge. For example with \textit{Athenaeum Place}, no extracted information mentions \textit{Athena} but surprisingly, the generator correctly infers the correct answer from its own knowledge. This capacity of generalization can be improved with a better quality of the top-10 items provided in the dynamic prompt. In Figure~\ref{fig:loss_modules_IR_graph}, we observe that the generator loses less information after the ranker if the top-$k_{\textrm{ranker}}$ has good scores on \textit{geo\_vic} in C2a~[\symbCTwoA~], which compensates lower results after the ranker: through in-context learning, the generator tends to favor answers related to Victoria or Australia. However, there is still a compromise between spatial or semantic scores, as shown in Figure~\ref{fig:resGenerator}.

\section{Discussion and conclusions}

Our experiment underlines the important differences between spatial proximity and semantic proximity. In our GIR architecture, later modules aim to counterbalance weaknesses of earlier ones. By exploiting grounded knowledge with different approaches, the final output is consolidated. In summary, the architecture aims to support the principle that \textit{maps still speak louder than words}. 

However, our architecture exhibits two stubborn weaknesses. First, spatial information is under-used; and second, losses of information propagate through our system. These weaknesses are particularly impactful for retrieval of long-tail information, such as the origin of place names. Spatially proximal information might be semantically distal in the language models. Other preliminary findings are also suggested by our results. 


\paragraph*{Co-dependencies with structured graphs instead of texts.} In our process, we treat each candidate independently in the ranking and we do not consider that each candidate might mutually contribute to better understand each other. An interesting direction is to develop multimodal information retrieval, that combines texts and spatial knowledge graphs \cite{lanSurveyComplexKnowledge2021}. Their graph representation is able to create dependencies in a corpus in the ranker module while offering low-dimension representations for fast rankings.


\paragraph*{Qualitative spatial reasoning.} Disambiguation is a key factor to improve geographic information retrieval, particularly to help to associate a footprint with the mentions ``near Collins Street'' or ``arrived in Australia.'' Recent works try to tackle this intrinsic nature of spatial information \cite{beydokhtiIntegratingLargeLanguage2024} in language models. Nevertheless, the domain is still an open problem.

\paragraph*{Further development.} In this work, the results can be extended at larger scales, notably other cities in Australia or in France for example where the ambiguity behind a place name might be limited. In this chapter, we focused on a task that requires high resources in terms of annotations for the evaluation. In a first step, the creation of a gold dataset between a place name and its origin and, in a second step, the annotation of each result in the top-10 and top-1.

\bigskip 

Despite being encapsulated in texts, spatial containment relations are better captured within hierarchies. In this aim, the high level of representation conveyed by knowledge graphs is more promising than prosaic texts. Techniques commonly applicable for natural language processing systems partially fail with spatial information: indeed, geographic information retrieval needs to know which footprints are impacted rather than which words are.

\section*{Acknowledgments} This project was undertaken with the assistance of computing resources from RACE (RMIT AWS Cloud Supercomputing). We also acknowledge the long-term work of Geographic Names Victoria with its past and current members. 

We also thank the two anonymous reviewers whose comments contributed to this chapter.

\bibliography{Bibliography.bib}

\end{document}